\documentstyle[epsfig,aps]{revtex}
\begin{document}
\draft
\title{Motional effects of single trapped atomic/ionic qubit}
\author{L. You}
\address{School of Physics, Georgia Institute of Technology,
Atlanta, Georgia 30332-0430, USA}

\date{\today}
\maketitle
\begin{abstract}

We investigate theoretical decoherence effects of the
motional degrees of freedom of a single trapped atomic/ionic
electronically coded qubit.
For single bit rotations from a resonant running wave
laser field excitation, we found the achievable
fidelity to be determined by a single parameter
characterized by the motional states.
Our quantitative results provide a useful realistic view
for current experimental efforts in quantum information and computing.

\end{abstract}
\pacs{03.67.Lx, 89.70.+c, 32.80.-t}

\narrowtext

Since the pioneering work by Shor \cite{Shor} on efficient prime factorization
with a quantum computer in 1994, we have witnessed an explosive growth of
interests in quantum information and computing. Although still largely a
theoretical field, solid progress in experimental efforts have been made
within the last few years. Most notably, pure state based
quantum gate implementation \cite{nist}, demonstrations of
quantum teleportation \cite{tele}, and GHZ state synthesis \cite{anton}
have stimulated more vigorous experimental efforts.

Two of the most interesting proposals for potentially large scale
quantum computing were suggested by the same
Innsbruck group, based on trapped ions \cite{Cirac}
and cavity QED with atoms \cite{Cirac2}. Various
implementations of these ideas are actively pursued in many
experiments around the world.
In their original analysis as presented in \cite{Cirac,Cirac2},
individual qubits are coded in electronic degrees states of atoms/ions,
and coherent evolution of the system state
requires the qubits to be in selected motional pure states.
Experimentally, one needs to attain the
strong binding limit \cite{side} and cooling to
the motional ground state is also required \cite{monroe}.
As is well known, these limits are difficult to maintain
due to various decoherence processes \cite{bf}
which heat up the motional degrees of freedom.
Further more, maintaining motional ground state becomes
problematic when strong confinement is not satisfied.
Several ideas were proposed recently for computing with
`hot' qubits \cite{Cirac3}.

This paper attempts to provide quantitative answers to
motional effects (ME) on electronically encoded quantum states \cite{Milburn}.
It is the first step towards a thorough investigation of
the ME. The paper is organized as follows.
First our model is presented.
We then discuss analytically the decoherence of the ME
for an unknown electronic encoded qubit.
Finally we present numerical results to support our understanding.
In forthcoming papers, we will study the decoherence
due to ME on multi-qubit entanglement creation,
e.g, the effect on the conditional logic
operation CNOT.

We consider a single harmonically bound two state atom
described by the Hamiltonian \cite{jas,zeng,vogel}
\begin{eqnarray}
H&&=\sum_{\vec n}\hbar(n_x\omega^g_x+
n_y\omega^g_y+n_z\omega^g_z)|g,\vec n\rangle\!\langle g,\vec n|\nonumber\\
&&+\sum_{\vec m}\hbar(\omega_{eg}+m_x\omega^e_x+
m_y\omega^e_y+m_z\omega^e_z)|e,\vec m\rangle\!\langle e,\vec m|\nonumber\\
&&+{1\over 2}\hbar\Omega_L e^{i\omega_L t}\sum_{\vec n,\vec m}
\eta_{\vec n\vec m}(\vec k_L)|g,\vec n\rangle\!\langle e, \vec m|
+h.c.\ .
\label{H_1}
\end{eqnarray}
$|g,\vec n\rangle=|g\rangle|\vec n\rangle_g$
($|e,\vec m\rangle=|e\rangle|\vec m\rangle_e$) denotes
number state in the ground (excited) trap with
frequencies are $\omega^g_{i=x,y,z}$ ($\omega^e_i$).
$\omega_{eg}$ is the electronic transition frequency.
$\Omega_L$ is the Rabi
frequency of the plane wave laser field.
The motional dipole moments are the
familiar Franck-Condon factor \cite{you}
\begin{equation}
\eta_{\vec n\vec m}({\vec k}_L)= \langle{g},{\vec n}|e^{-i{\vec k}_L\cdot{\vec R}
} |{e},{\vec m}\rangle.
\label{fc}
\end{equation}

Radiative coupling to the vacuum reservoir of the atom
will not be included here as the effect of the
resulting spontaneous emission on the qubit decoherence
has been studied and is well understood \cite{Cirac,Cirac2,Cirac3,walls}.
Our model can also be viewed as between two ground states
in a three level $\Lambda$-type off resonant Raman system.
In such a case,
$\Omega_L\sim\Omega_P\Omega_S^*/\delta_L$ are the two photon
effective Rabi frequency and $\vec k_L=\vec k_P-\vec k_S$.
The indices $P$ and $S$
denote the dipole connected pump and stokes transitions,
and $\delta_L$ is the (large) detuning
from the eliminated far off-resonant excited state.

Physical models similar to Eq. (\ref{H_1}) have been
studied under different context before \cite{Cirac,side,zeng,vogel,gorigi},
usually, within the Lamb-Dicke limit (LDL) when
ME become considerably simplified. In terms of the trap width
$a_i^{g,e}=\sqrt{\hbar/2M\omega_i^{g,e}}$,
the LDL corresponds to $k_La_i^{g,e}\ll 1$.
This requires the atom/ion to be confined less than the
wavelength $a_i^{g,e}\ll\lambda_L$ and is equivalent to
require $\hbar\omega_i^g$ to be much larger
than the recoil energy $E_R=\hbar^2k_L^2/2M\ll \hbar\omega_i^{e,g}$.
For an effective two state system reduced from a
near resonant three level $\Lambda$-type configuration,
LDL is easily satisfied with co-propagating
pump and Stokes fields when $\vec k_P\sim\vec k_S$.

In this study we investigate ME for general cases not in the LDL.
Such studies will provide
much needed theoretical clarification as trapped atoms/ions are
among the ``hottest'' qubit candidates in many experimental efforts.
We note Franck-Condon factors (\ref{fc}) satisfy
\begin{eqnarray}
&&\sum_{\vec n}\big[\eta_{\vec n\vec m}({\vec k}_L)\big]^{*} \eta_{\vec n%
\vec m^{\prime}}({\vec k}_L)=\delta_{\vec m\vec m^{\prime}}, \nonumber\\
&&\sum_{\vec m}\big[\eta_{\vec n\vec m}({\vec k}_L)\big]^{*} \eta_{\vec n%
^{\prime}\vec m}({\vec k}_L)=\delta_{\vec n\vec n^{\prime}},
\end{eqnarray}
which allows the introduction of a complete and orthonormal basis \cite{you}
\begin{eqnarray}
|\vec n\rangle_p = \sum_{\vec n'}\eta_{\vec n\vec n'}^*({\vec k}_L)|\vec n'\rangle_g.
\label{wp}
\end{eqnarray}
Physically it corresponds to the motional wave packet
of a photon absorption from ground state $|\vec n\rangle_g$.
Mathematically, it is the number
coherent state basis $|\alpha,n\rangle=D(\alpha)|n\rangle$, with
the displacement operator $D(\alpha)=e^{\alpha b^+-\alpha^* b}$.
We note $R_i=a_i^g(b_i^{g+}+b_i^g)$
with $b_i^g$ ($b_i^{g+}$) the annihilation (creation)
operator for$|n_i\rangle_g$, therefore Eq. (\ref{wp}) can be
rewritten as $|n_x\rangle_p =|ik_La_x^g,n_x\rangle_g$, exactly
representing wave-packets
corresponding to excitation from different ground trapping
states. As will become
clear later, the coherent Rabi coupling between
the ground and excited state manifolds can
also be decomposed into paired sets
$\{|g\rangle|\vec n\rangle_g,|e\rangle|\vec n\rangle_p\}$.

With the inverse relation
\begin{eqnarray}
|\vec m\rangle_e=\sum_{\vec n}\eta_{\vec n\vec m}({\vec k}_L)|\vec n\rangle_p,
\end{eqnarray}
we can transform Eq. (\ref{H_1}) into the $|\vec n\rangle_p$ basis.
Denote $|e_p,\vec n\rangle=|e\rangle|\vec n\rangle_p$, we obtain
\begin{eqnarray}
\sum_{\vec m}\hbar\omega_{eg}|e,\vec m\rangle\!\langle e,\vec m|
&&=\sum_{\vec n}\hbar\omega_{eg}|e_p,\vec n\rangle\!\langle e_p,\vec n|,\nonumber\\
\sum_{\vec m,i} m_i\hbar\omega^e_i
|\vec m\rangle_e\!\ _e\langle\vec m|
=&&\sum_{\vec n\vec n'}
(E_{\vec n\vec n'}^D+E_{\vec n\vec n'}^O)
|\vec n\rangle_p\!\ _p\langle\vec n'|.
\end{eqnarray}
We found $E_{\vec n\vec n'}^D$ terms couple
nearest neighbors, i.e.
states with $n_i=n_i'\pm 1$ ($n_{j\ne i}=n'_j$),
while $E_{\vec n\vec n'}^O$ terms couple states with
$\vec n=\vec n'$
and $n_i=n_i'\pm 2$ ($n_{j\ne i}=n'_j$).
$E_{\vec n\vec n'}^O$ terms become diagonal
along $i-$axis whenever $\delta\omega_i^2\neq 0$.

We will focus on the case
$\omega_i^g=\omega_i^e$ ($i=x,y,z$) in the
present paper. This is typical for the ion trap system
and can also be arranged for optical dipole traps \cite{jeff}.
For the plane wave excitation along the x-axis,
the ME along the $y$ and $z$ directions are unperturbed.
Our Hamiltonian Eq. (\ref{H_1}) can simplifies to
a one dimensional model
\begin{eqnarray}
H&&=\sum_{n_x}n_x\hbar\omega^g_x|g,n_x\rangle\!\langle g,n_x|\nonumber\\
&&+\sum_{n_x}\hbar(n_x\omega^g_x-\Delta_L)
|e_p,n_x\rangle\!\langle e_p,n_x|\nonumber\\
&&+i(k_La_x)\sum_{n_x}\hbar\omega^{g}_x \sqrt{n_x+1}
|e_p,n_x+1\rangle\!\langle e_p,n_x|+h.c.\nonumber\\
&&+{1\over 2}\hbar\Omega_L
\sum_{n_x}|g,n_x\rangle\!\langle e_p,n_x|+h.c. ,
\label{hm}
\end{eqnarray}
where transformation to the interaction picture by
\begin{eqnarray}
U(t)=\exp\left(-i\hbar\omega_Lt\sum_{n_x}|e_p,n_x\rangle\!\langle e_p,n_x|\right)
\end{eqnarray}
has also been made. The detuning is
$\hbar\Delta_L
=\hbar\omega_L-\hbar\omega_{eg}-{\hbar^2k_L^2\over 2M}$,
including the recoil shift.

\begin{figure}[t]
\centerline{\epsfig{file=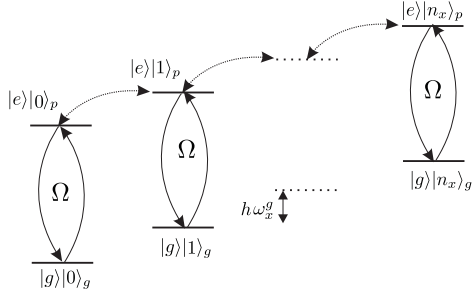,width=2.5in}\\[12pt]}
\caption{The paired ladders of a trapped two state atom/ion.
Solid curve arrow heads denote Rabi oscillations between
paired states, while the dotted curve arrow heads denote
nearest neighbor motional coupling.}
\label{fig1}
\end{figure}

This Hamiltonian can be graphically illustrated as
in Figure \ref{fig1}. The paired
ladder structure resembles the familiar
motional state ladders in an ion trap \cite{side}.
However, in Fig. \ref{fig1},
the excited states are the wave-packet basis states Eq. (\ref{wp}).
The nearest neighbor coupling is not
due to the LDL approximation.
Denote $H=H_0+H_1$ with $H_1$ the nearest neighbor coupling term
in the excited state [the third line of Eq. (\ref{hm})], $H_0$
becomes
\begin{eqnarray}
H_0=\hbar\left (\begin{array}{cccccccccc}
\cdot & \cdot & & & & \\
\cdot & \cdot & & & & \\
& & n_x\omega_x^g & {\Omega_L\over 2} & & \\
& & {\Omega_L\over 2} & n_x\omega_x^g-\Delta_L & & \\
& & & & \cdot & \cdot \\
& & & & \cdot & \cdot \\
\end{array} \right),
\end{eqnarray}
where each $2\times2$ block describes
the Rabi oscillation
between paired states $\{|g\rangle|n_x\rangle_g,|e\rangle|n_x\rangle_p\}$
with {\it exactly the same} Rabi
frequency $\Omega=\sqrt{\Omega_L^2+\Delta_L^2}$.
There are no differential detunings between different pairs either.
With wave function coefficients $\{C_{n_x}^g,C_{n_x}^e\}$,
the $2\times2$ oscillation is described by
\begin{eqnarray}
C_{n_x}^e(\tau) &&=e^{-in_x\omega_x^g\tau}
\left[C_{n_x}^e(0)\cos\theta-iC_{n_x}^g(0)\sin\theta\right],\nonumber\\
C_{n_x}^g(\tau) &&=e^{-in_x\omega_x^g\tau}
\left[C_{n_x}^g(0)\cos\theta-iC_{n_x}^e(0)\sin\theta\right],
\label{twou}
\end{eqnarray}
when $\Delta_L=0$. The pulse area is
$\theta(\tau)={1\over 2}\int_{0}^\tau\Omega(t) dt$.
$H_0$ describes the coherent evolution
between paired states with a time scale given $\Omega$.
$H_1$, on the other hand couples nearest neighbors
of excited motional wave pack states and can causes decoherence
of an electronically coded qubit. Its time scale is
determined by several factors including the
trap frequency $\omega_x^g$, Lamb Dicke parameter $k_La_x^g$,
and the highest motional state number $n_x^{\rm max}$.
Assuming an electronically coded unknown qubit
\begin{eqnarray}
|\psi(0)\rangle&&=\alpha |g\rangle +\beta |e\rangle,
\label{qb}
\end{eqnarray}
(normalization $|\alpha|^2+|\beta|^2=1$),
an arbitrary single bit
rotation is achieved through a multiplication of \cite{barenco}
\begin{eqnarray}
e^{i\delta}\left( \begin{array}{cc}
e^{i\gamma_1} & 0\\
0 & e^{-i\gamma_1} \end{array}\right)
\left( \begin{array}{cc}
\cos\theta & \sin\theta\\
-\sin\theta & \cos\theta \end{array}\right)
\left( \begin{array}{cc}
e^{i\gamma_2} & 0\\
0 & e^{-i\gamma_2} \end{array}\right).\nonumber
\end{eqnarray}
$\delta$, $\gamma_1$, $\gamma_2$, and $\phi$ are parameters.
With resonant Rabi coupling, our solution Eq. (\ref{twou})
achieves the important
$\theta$ rotation corresponds to
$e^{2i\gamma_1}=-i$, and $e^{2i\gamma_2}=i$. Ideally
one hopes to arrive at the target state
\begin{eqnarray}
|\psi(\tau)\rangle_T&&=
(\alpha\cos\theta-i\beta\sin\theta) |g\rangle
+(\beta\cos\theta-i\alpha\sin\theta) |e\rangle,\nonumber
\end{eqnarray}
with the density matrix,
\begin{eqnarray}
\rho_T(\tau)&&=I_{g}^T|g\rangle\!\langle g|
+(1-I_{g}^T)|e\rangle\!\langle e|
+(I_{ge}^T|g\rangle\!\langle e|+h.c.),
\label{rhot}
\end{eqnarray}
where
\begin{eqnarray}
I_{g}^T&&=|\alpha|^2\cos^2\theta+|\beta|^2\sin^2\theta
+i(\alpha\beta^*-c.c.)\sin\theta\cos\theta,\nonumber\\
I_{ge}^T&&=\alpha\beta^*\cos^2\theta+\beta\alpha^*\sin^2\theta
+i(|\alpha|^2-|\beta|^2)\sin\theta\cos\theta.\nonumber
\end{eqnarray}

Due to ME the electronic qubit Eq. (\ref{qb})
does not remain decoupled from the motional degrees of freedom.
In general, it evolves within the
much larger Hilbert space containing motional states.
We now study two concrete examples of
decoherence assuming the initial qubit
in the enlarged Hilbert space reproduces the
density matrix Eq. (\ref{rhot}), i.e. resembles
a perfect qubit to the innocent bystanders unaware
of the motional degrees of freedom.

First, we consider
\begin{eqnarray}
|\psi(0)\rangle_{\rm tot}
&&=(\alpha |g\rangle + \beta|e\rangle)\otimes |\psi(0)\rangle_{\rm cm},
\label{psi1}
\end{eqnarray}
with an initial pure motional state
$|\psi(0)\rangle_{\rm cm}=\sum_{n_x}c_{n_x}|n_x\rangle_g$
($\sum_{n_x}|c_{n_x}|^2=1$).
Upon tracing the motional degrees of freedom,
$\rho(0)=|\psi(0)\rangle\!\langle\psi(0)|$ is correctly
reproduced. By rewriting $|\psi(0)\rangle_{\rm tot}$
as
$$
\sum_{n_x}(c_{n_x}\alpha |g\rangle|n_x\rangle_g
+\sum_{n_x'}\eta_{n_x n_x'}c_{n_x'}\beta |e\rangle |n_x\rangle_p),
$$
we can analytically
evolve this state with $H_0$ to obtain
\begin{eqnarray}
\rho(\tau)&&=I_{g}|g\rangle\!\langle g|
+(1-I_g)|e\rangle\!\langle e|
+(I_{ge}|g\rangle\!\langle e|+h.c.),
\label{rho1}
\end{eqnarray}
with
\begin{eqnarray}
I_{g} &&=|\alpha|^2\cos^2\theta+|\beta|^2\sin^2\theta
+i(\alpha\beta^*\eta^*-c.c.)\cos\theta\sin\theta,\nonumber\\
I_{ge} &&=i(|\alpha|^2-|\beta|^2){1\over 2}\sin2\theta
\sum_{n_x,q_x}e^{-i(n_x-q_x)\omega_x^g\tau}
c_{n_x}c_{q_x}^*\eta_{q_xn_x}\nonumber\\
&&+\alpha\beta^*\cos^2\theta
\sum_{n_x,q_x}e^{-i(n_x-q_x)\omega_x^g\tau}
c_{n_x} \sum_{q_x'}\eta^*_{q_x q_x'}c_{q_x'}^*\eta_{q_xn_x} \nonumber\\
&&+\alpha^*\beta\sin^2\theta
\sum_{n_x,q_x}e^{-i(n_x-q_x)\omega_x^g\tau}
\sum_{n_x'}\eta_{n_x n_x'}c_{n_x'}c_{q_x}^*\eta_{q_xn_x},\nonumber
\end{eqnarray}
where we have defined the parameter
\begin{eqnarray}
\eta(k_L)=\sum_{n_x,n_x'}c_{n_x}^*\eta_{n_x n_x'}c_{n_x'}.
\end{eqnarray}
We see the evolution by Eq. (\ref{twou})
will in general not reproduce the intended
density matrix Eq. (\ref{rhot}) because of $\eta$.
$H_1$ term is the other reason for incomplete control
although its effects (when $\Omega_L\gg\omega_x^g$)
can be minimized by employing a fast pulse
with $\omega_x^g\tau\ll 1$.
Within such a limit, or when $\omega_x^g\tau=2\pi$,
we obtain
\begin{eqnarray}
I_{ge} &&=i(|\alpha|^2-|\beta|^2)\sin\theta\cos\theta\,\eta(k_L)\nonumber\\
&&+\alpha\beta^*\cos^2\theta+\alpha^*\beta\sin^2\theta\,\eta(2k_L).
\end{eqnarray}
The necessarily condition for attending
a perfect fidelity of the single bit
rotation is then $\eta(k_L)\equiv 1$, which can be
approximately satisfied in the LDL when $k_La_x^g\ll$
or in the $\Lambda$-type Raman systems with
co-propagating pump and Stokes fields.
As a second example, we consider the case of a
thermal motional state
\begin{eqnarray}
\rho_{\rm tot}(0) &&=|\psi(0)\rangle\!\langle\psi(0)|\otimes\rho_{\rm cm}(0),\nonumber\\
\rho_{\rm cm}(0) &&= \sum_{n_x}\rho^{\rm cm}_{n_x} |n_x\rangle\!\langle n_x|,\nonumber\\
\rho^{\rm cm}_{n_x} &&
=(1-e^{-\hbar\omega_x^g/k_BT})e^{-n_x\hbar\omega_x^g/k_BT}.
\end{eqnarray}
This is the limiting case of an ensemble average
of Eq. (\ref{psi1})
with
$c_{n_x}=\sqrt{\rho_{n_x}^{\rm cm}}\,e^{-i\phi_{n_x}}$
and
$\phi_{n_x}$ a uniform random number $\in[0,2\pi)$.
The dynamics due to $H_0$ can be evolved analytically
and the same density matrix Eq. (\ref{rho1}) is
obtained. After averaging over ${\{\phi_n\}}$, we obtain
\begin{eqnarray}
\langle \eta(k_L)\rangle_{\{\phi_n\}}
 &&=\sum_{n_x} \rho_{n_x}^{\rm cm}  \eta_{n_xn_x}(k_L)\nonumber\\
&&=\exp\left[-{1\over 2}(k_La_x)^2\coth\left({1\over 2}{\hbar\omega_x^g\over k_BT}\right)\right].
\end{eqnarray}
In the low temperature limit when
$k_BT< \hbar\omega_x^g$, $\eta$ becomes $1$ as
long as LDL $k_La_x^g\ll 1$ is satisfied.
At high temperatures when $k_BT\gg\hbar\omega_x^g$,
$(k_La_x^g)^2\ll {\hbar\omega_x^g/k_BT}$ needs to satisfied
for $\eta$ close to $1$.

We now discuss the numerical solutions.
Expand the total wave-function as
\begin{eqnarray}
|\psi(t)\rangle_{\rm tot}=\sum_{n_x}[c_{n_x}^g(t)|g\rangle|n_x\rangle_g
+c_{n_x}^e(t)|e\rangle|n_x\rangle_p],
\end{eqnarray}
we have solved the Schr\"odinger equation
including both $H_0$ and $H_1$.
The transformation Eq. $(\ref{wp})$ greatly
reduces the motional Hilbert space dimension.
The perceived fidelity
for the electronic coded qubit Eq. (\ref{qb})
under transformation Eq. (\ref{rhot}) is
\begin{eqnarray}
{\cal F} &&=\rm Tr[\rho_T(\tau)\rho(\tau)].
\end{eqnarray}

\begin{figure}[t]
\centerline{\epsfig{file=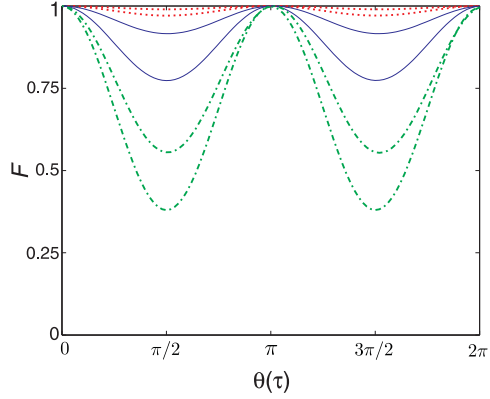,width=2.5in}\\[12pt]}
\caption{The fidelity for single bit rotations with
$k_La_x^g=0.1$ (dashed lines), $0.3$ (solid lines),
and $1.0$ (dot dashed lines). The higher and lower fidelity
sets are for initial motional state $c_{n_x}=\delta_{n_x0}$ and
$c_{n_x}=(2\,\delta_{n_x0}
+\sqrt{2}\,\delta_{n_x1}+\delta_{n_x2})/\sqrt{7}$ respectively.
$\Omega_L=100$ $(\omega_x^g)$.}
\label{fig2}
\end{figure}

We take $\alpha=\beta=1/\sqrt{2}$ as an example to illustrate
our numerical results since similar/better fidelities
are obtained with other choices. In Figure \ref{fig2}
we compare fidelities under arbitrary $\theta(\tau)$
rotations for two different pure states.
Acceptable fidelities are obtained only for $k_La_x^g\le 0.3$ .
In general larger $\Omega_L/\omega_x^g$ ratios
also improved fidelity although it saturates
around $\Omega_L/\omega_x^g \sim 100$.
Noticeable improvements are also recorded for
narrower distributions in $|c_n|^2$, e.g. in Fig. \ref{fig2}
initial state with $c_{n_x}=\delta_{n_x 0}$
produced better fidelity. This is a direct reflection
of dephasing among different motional pair states
because of their different time scales from $H_0$
and $H_1$. The oscillatory behavior is due to dephasing
caused by the Rabi oscillation between motional paired
states. For comparison, we note that
meaningful single bit rotations need to achieve
a fidelity of $1/2$, the lower limit from a random (uncontrolled)
sampling of final states.

\begin{figure}[t]
\centerline{\epsfig{file=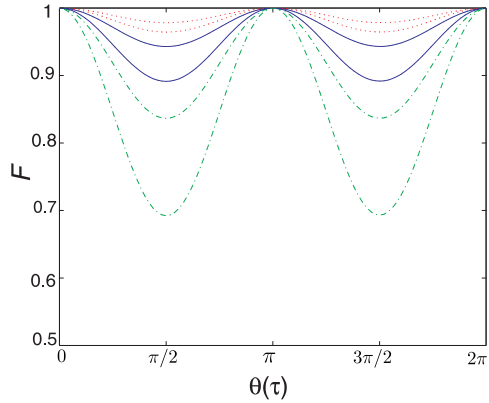,width=2.5in}\\[12pt]}
\caption{The fidelity for single bit rotations
with $k_BT/\hbar\omega_x^g=1$ (dashed lines), $3$ (solid lines),
and $10$ (dot dashed lines). The higher and lower fidelity
sets are for initial motioal thermal state $\rho_{n_x}^{\rm cm}$
and pure state $c_{n_x}=\sqrt{\rho_{n_x}^{\rm cm}}$ respectively.}
\label{fig3}
\end{figure}

Finally we compare with thermal states
for several different values of $k_BT/\hbar\omega_x^g$.
Surprisingly, we found the fidelity for an initial
motional thermal state is always higher than
its corresponding pure state. In the temperature
regime considered we found
acceptable results as long as LDL is maintained.
In Fig. \ref{fig3}, we have used
$k_La_x^g=0.1$ and $\Omega_L=100$ ($\omega_x^g$).

In conclusion, we have performed detailed
theoretical studies of the decoherence of an electronically
coded atom/ion qubit due to ME. By introducing a
wave packet basis in the excited state, we were able
to perform considerably cleaner analysis to simplify
the ME. We found that a single parameter $\eta$ measures
the achievable fidelity of arbitrary single bit rotations.
We performed numerical calculations which demonstrates our
understanding and provided quantitative limits
for experiments: the LDL is always required to
maintain a high fidelity for arbitrary single bit rotations.
We also found that a pure motional state is not necessarily
preferred although a qubit with an initial ground
motional state does give rise to the
highest recorded fidelity. In actual experimental
implementations, a large $\Omega_L$ is also needed
to assure negligible motional dephasing during $\tau$.
One can always wait for a period $2\pi/\omega_x^g$ for subsequent
single bit operations since the motional wave function then
always rephases to its initial state.
This study will also shed light on devising schemes for overcoming
ME decoherence and error corrections in
trapped atomic/ionic qubits.

We thank Dr. T. Uzer and Dr. M. S. Chapman for helpful discussions.
We also thank Dr. M\"{u}stecapl{\i}o\={g}lu
for supplying a Fortran
subroutine for evaluating $\eta_{n_xm_x}$. This work is supported by
the ONR grant No. 14-97-1-0633 and the ARO/NSA grant G-41-Z05.

%
%
%
%
\newpage

\end{document}